# Sensing of streptococcus mutans by microscopic imaging ellipsometry

**Mai Ibrahim Khaleel,**[a,b,c] **Yu-Da Chen,**[a,b,c] **Ching-Hang Chien,**[a,b,c] **Yia-Chung Chang**[a*]

[a] Research Center for Applied Sciences, Academia Sinica, Taipei, 11529, Taiwan
[b] Nano Science and Technology Program, Taiwan International Graduate Program, Academia Sinica and National Tsing Hua University, Taiwan
[c] Department of Engineering and System Science, National Tsing Hua University, Hsinchu, 30013, Taiwan

**Abstract**. Microscopic Imaging Ellipsometry is an optical technique that uses an objective and sensing procedure to measure the ellipsometric parameters $\Psi$ and $\Delta$ in the form of microscopic maps. This technique is well known for being non-invasive and label-free. Therefore it can be used to detect and characterize biological species without any impact. In this work MIE was used to measure the optical response of dried Streptococcus mutans cells on a glass substrate. The ellipsometric $\Psi$ and $\Delta$ maps were obtained with Optrel Multiskop system for specular reflection in the visible range ($\lambda$= 450nm -750nm). The $\Psi$ and $\Delta$ images at 500nm, 600nm, and 700nm were analyzed using three different theoretical models with single-bounce, two-bounce, and multi-bounce light paths to obtain the optical constants and height distribution. The obtained images of the optical constants show different aspects when comparing the single-bounce analysis with the two-bounce or multi-bounce analysis in detecting S. mutans samples. Furthermore, the height distributions estimated by two-bounce and multi-bounce analysis of S. mutans samples were in agreement with the thickness values measured by AFM, which implies that the two-bounce and multi-bounce analysis can provide information complementary to that obtained by single-bounce light path.

*Yia-Chung Chang, E-mail: yiachang@gate.sinica.edu.tw

## 1 Introduction

Spectroscopic ellipsometry (SE) is a popular metrology technique used to determine thin film thickness and its optical constants. The technique allows us to reconstruct a nanoscale resolution image of the material surface by fitting the measured ellipsometric parameters Psi ($\Psi$) and Delta ($\Delta$) with a suitable theoretical simulation.[1] Here $\Psi$ is a measure of the absolute value of the ratio of the complex Fresnel reflection coefficients (or reflectivities), $r_p$ and $r_s$, for p- and s-polarized light, and $\Delta$ is the phase difference between $r_p$ and $r_s$. Combing SE with microscopy leads to a powerful noncontact imaging technique known as Microscopic Imaging Ellipsometry (MIE).[2] MIE enables a large area imaging with a spatial resolution of 1µm and subnanometer-scale z-resolution.[2, 3] In addition to the sub-nanoscale thickness resolution, it does not require a vacuum ambient nor the labeling of molecules, as compared to other detection techniques that require labeling - usually fluorescent or enzymatic such as Fluorescence imaging.[4-6] The labeling itself in such techniques can be highly heterogeneous, leading to non-quantitative results.[6] MIE enables faster real-time measurements (in seconds) in comparison with other slow scanning techniques such as AFM that requires a few minutes to scan one area. This is especially useful for biological processes that occur within seconds rather than minutes.[7, 8]. These unique features show the versatility of this imaging technique which had increased its usage in sensing and analyzing biological systems.

One of the pioneering applications of imaging ellipsometry was to detect antigen-antibody interactions using single wavelength light source and fitting the measured data before and after interaction happens.[9] Other applications of imaging ellipsometry were reported, including studies of biological reaction in cancer cells,[10] detection of insulin-antibody binding on solid substrate,[11] detection of latent fingermarks,[12] as well as mapping of multilamellar films, hydrated membranes, and Fluid domains.[13]

Streptococcus mutans (S. mutans) is an aerobic, gram-positive bacterium with coccus shape found in the human oral cavity.[14, 15] It has been reported as the main etiological agent of dental caries and normal inhabitant of dental plaque.[16] S. mutans are known for its ability to synthesize extracellular polymeric substance (EPS), which works as a matrix holding microbial cells together to form a biofilm that attached to teeth surface. This matrix serves as the colony of bacteria and contributes to structure maintenance and antibiotic resistance of the biofilm.[17, 18] In the metabolism process S. mutans produces aggressive substances like acids, which cause a low pH environment in the mouth and lead to corrosion of enamel and the formation of cavities.[18, 19].

In recent years several research groups have developed methods for detecting and studying S. mutans in either laboratory settings or field trials. To our knowledge, there have been no attempts to detect S. mutans cells in the visible region using MIE. In this work, we performed MIE measurements of S. mutans cells grown on a glass substrate and dried in air. We assumed a structure consisting of air/bacterial cells/glass substrate to study the dielectric properties and height distribution. The measured $\Psi$ and $\Delta$ maps were fitted using three different optical models to detect bacterial cells and the surrounding dried culturing medium. In the first model, we considered only single-bounce light path with the reflection from the interface between air and bacterial cells. In the second model, we considered the coherent superposition of the reflection from the interface between glass substrate and sample and the first reflection and called this "two-bounce" model. In the third model, we included all reflections in the structure and called it "multi-bounce" model. AFM Veeco Innova,[20] tapping mode was used in this study to map the surface topography of S.mutans cells and culturing medium. The observed height distributions of cells and culturing medium on glass substrate by AFM were compared to the deduced values of height distribution by two-bounce and multi-bounce analysis. The glass substrate was chosen to grow S.mutans because of two main reasons. The first one is that a series of studies have been done by using a glass substrate to study and reveal many details about S.mutans adherence and other oral bacteria.[21] The second reason is that glass substrate is easy to calibrate which makes the analyses simpler.

## 2  Materials and Methods

*2.1 Cell Cultures*

An S. mutans strain (ATCC 25175) was purchased from Creative Life Sciences (CMP)-Taipei-Taiwan.[22]   S. mutans was incubated in brain-heart infusion broth (BHI) purchased from Creative Life Sciences (CMP)-Taipei-Taiwan,[22] shaken at 37 ºC overnight. Another BHI solution was prepared to grow bacterial cells on a glass substrate. The solution consists of 7.4g BHI, 20g sucrose, and 200 ml Millipore water.[23] Sucrose was added in order to enhance the rate of cells growth on a glass substrate.

*2.2  Cell Sample Preparation*

Glass slides (BRAND GMBH + CO KG, Wertheim, Germany),[24] were cut into 1×1 cm$^2$ pieces. They were cleaned through sonication in acetone for 30 minutes, then for 20 minutes in isopropanol solution at room temperature. After that glass pieces were washed and sonicated in Millipore water for 10 minutes and finally dried in an oven at 60 Cº. S. mutans was cultivated on glass substrates as described by Hu et al.[23] The cleaned glass substrates were immersed into a Petri dish containing 10 ml of freshly prepared BHI medium with sucrose. Then 2.5 ml of incubated S. mutans cells in BHI medium overnight was added to the Petri dish contacting the immersed substrates, sealed by Parafilm, and incubated for 72 h at 37 ºC. The glass samples were then washed three times with Millipore water and placed in a new Petri dish to dehydrate naturally for one day, then measured by MIE and AFM. Bare glass substrates were kept in culturing medium (BHI and sucrose) without S.mutans inoculated were used in reference experiments.

*2.3 Multiskop Imaging Ellipsometry*

The ellipsometric parameters $\Psi$ and $\Delta$ of S. mutans cells on glass substrate were measured using an Optrel Multiskop system,[25] as shown in Fig.1 which is modified to a rotating compensator ellipsometry. Converting the system from null measurements (which requires rotating both the polarizer and analyzer) to RCE allowed us to avoid the image-shift problem caused by the rotating analyzer (a Glan–Taylor prism [26]), and to a faster extraction of the ellipsometric image.[1]

The setup of our MIE consists of laser arm, sample stage, and detector arm. The light beam comes from a Fianium WhiteLase Short Wavelength Supercontinuum laser that covers wavelengths from 400nm to 1000nm with power > 300mW,[27] then passes through a polarizer fixed at 45º and a rotating compensator. To enhance the intensity of the incident light a 10x lens was inserted before the light hits the spot of interest on the sample. The reflected or scattered light from the sample then goes through an 80x objective lens,[28] that makes our system capable of probing an area as small as 1µm × 1µm. The light beam then passes through an analyzer mounted at 90º with respect to the polarizer, and finally collected by a charge-coupled device (CCD) camera (Pixelfly qe), [29] with a pixel size equal to 164 nm. Measurements of ellipsometric parameters in this study were done using specular mode. In this mode, both the laser arm and detector arm are rotated in opposite directions with the use of a goniometer. The data collected by the CCD are the reflected light beam from the sample which are analyzed to determine the ellipsometric parameters $\Psi$ and $\Delta$. In RCE, the reflected light intensity as a function of compensator angle $\phi_c$ is given by Ref. [10]

$$I(\phi_c) = A_0 + A_2 \cos(2\phi_c) + B_2 \sin(2\phi_c) + A_4 \cos(4\phi_c) + B_4 \sin(4\phi_c). \tag{1}$$

We take measurements of $I(\phi_c)$ for 36 values of $\phi_c$ starting from 5° and ending at 355° with 10° spacing. These data allow us to determine the Fourier coefficients in Eq. (1), which can be used to evaluate $\Psi$ and $\Delta$.

*2.4 Theory and Numerical Calculations*

The phase difference between $r_p$ and $r_s$ (where $r_p$ and $r_s$ are reflectivity for p- and s-polarized light beam) and the ratio of their amplitudes can be quantified by the ellipsometric measurements as $\Delta$ and $\tan\Psi$, respectively. The analysis of RCE signals was done to determine $\Psi$ and $\Delta$ in terms of Fourier coefficients of the output light intensity as a function of the compensator angle.[10]

The fundamental equation of ellipsometry is then written as

$$\rho = \tan\Psi \; e^{i\Delta} = \frac{r_p}{r_s}, \tag{2}$$

where $r_p$ and $r_s$ are complex reflectivities for $p$ and $s$ polarizations. For the sample under consideration, the reflectivities can be calculated from Fig. 2 using three different models. The light detected by an image sensor is composed of light paths after a single bounce, two bounces, or multiple bounces.[30] For the single-bounce case, the coherent superposition of directly reflected light after a single reflection event with the sample is detected, while in the two- and multi-bounce cases the light is allowed to hit the interface between the sample and the substrate before reaching the sensor.[31]

In traditional methods of shape and reflection estimations such as time of flight and photonic mixer devices for opaque objects the single-bounce measurement is commonly used, while the multi-bounce effect is ignored since it is considered as a source of noise.[31] Recent studies of shape and reflection estimations have shown that two-bounce and multi-bounce light paths can provide information not achievable by using only single-bounce analysis. The two-bounce light paths were used for recovery of shape and reflection estimations,[31] while multi-bounce light paths have been used to characterize flat and transparent thin films using spectroscopic rotating polarizer-analyzer ellipsometer (RPAE).[32] To fully characterize the sample under consideration in this paper we incorporated analyses of single-bounce, two-bounce, and multi-bounce models for the data obtained by using RCE. The derivation of formulas relevant to the three models to be used below is given in Appendix A.

*2.4.1 Single-Bounce Light Path*

The single-bounce light path was used to detect bumpy areas, including S. mutans cells and the culturing medium on glass substrate. Since each grid (with 5 pixel average) of bumpy area has a curved surface and size of 820nm which is comparable to the wavelength, it is assumed that light going through the two-bounce and multi-bounce paths can be neglected. [See Fig. 2(a)] Therefore, light scattered from bumpy areas may be better described by the coherent sum of single-bounce light rays. By calculating $r_p$ and $r_s$ we can determine the refractive index $n_1$ and the extinction coefficient $k_1$ of bumpy areas according to the following equation.

$$\varepsilon_1 = (n_1 + ik_1)^2 = \frac{(1-\rho)^2}{(1+\rho)^2} \sin\theta_0^2 \tan\theta_0^2 + \sin\theta_0^2, \tag{3}$$

where $\varepsilon_1$ is the complex refractive index of bumpy area.

*2.4.2 Two-Bounce Light Path*

The model with two-bounce light path is used to detect to detect S. mutans cells and the dried culturing medium which represents a mixture of rough and nearly flat areas on the glass substrate. For the two-bounce light path depicted in Fig. 2(b) the reflectivity is given by:

$$r_j = r_{01j} + \left(1 - r_{01j}^2\right) r_{12j}\, e^{2i\beta}\ ,\tag{4}$$

where *j* stands for *s*-polarization and *p*-polarization, and $\beta$ represents the phase factor for light going through the film given by the equation below:

$$\beta = \frac{2\pi h n_1 \cos\theta_1}{\lambda}\ ,\tag{5}$$

where *h* and $n_1$ are the thickness and refractive index of the sample. Assuming the sample does not absorb light in the visible region (*i.e.*, $k_1=0$), then we have

$$\ln|\gamma| = \ln\left|e^{2i\beta}\right| = \ln\left|\frac{r_{01p} - \rho r_{01s}}{\rho\left(1 - r_{01s}^2\right)r_{12s} - \left(1 - r_{01p}^2\right)r_{12p}}\right| = 0\ ,\tag{6}$$

An in-house MATLAB program,[33] was used to solve Eq. (6) and find the refractive index and thickness of the cells and medium.

*2.4.3 Multi-Bounce Light Paths*

The model with multi-bounce light paths is also used to analyze the data. Based on the results of Fig. 4(a) and (b) we expect this model will give similar results as two-bounce model (for our case with AOI = 52° and height = 80 nm and 350nm), since the higher order terms in the multi-bounce model appear to be small. The reflectivity of Multi-bounce model is given by:

$$r_j = \frac{r_{01j} + r_{12j}\, e^{2i\beta}}{1 + r_{01j} r_{12j}\, e^{2i\beta}}\ ,\tag{7}$$

where $r_{01j}$ and $r_{12j}$ for *s*-polarization and *p*-polarization have been derived in Ref.[32]. Assuming that $k_1 = 0$ in the visible range, and the final solution is written as:

$$\ln|\gamma| = 0\ ,\tag{8}$$

where $\gamma$ is given in Eq. (A11) of Appendix A. Our in-house MATLAB program was also used to solve Eq. (8) and find the refractive index $n_1$ of the sample. Given $n_1$, the thickness of the sample can be obtained from Eq. (5) in terms of $\beta$, $n_1$, and $\cos \theta_1$ as:

$$h = \frac{\beta \lambda}{2 \pi n_1 \cos \theta_1} ,  \qquad (9)$$

## 3 Results and Discussion

For MIE measurements three samples of S.mutans were investigated, while only the results for two samples are reported here, since the third sample gives qualitatively the same results. A carbon tape was attached to the backside of the glass substrate of these samples in order to eliminate backside reflections of the glass substrate in our measurements. For AFM measurements, two samples were investigated using tapping mode to map the surface morphology of cells and culturing medium. All samples investigated by AFM and MIE were prepared under the same conditions.

*3.1 AFM Results*

AFM tapping mode was used to morphologically characterize S. mutans cells cultured on a glass substrate for 72 hours. All images were obtained with a commercial AFM (Veeco Innova) of ambient conditions. The OMCLO-AC160TS, R3 tip was used for tapping mode images. [34] Fig. 3 shows AFM images of S. mutans cells and heights distribution on a glass substrate with spatial resolution of 78nm/pixel.

The image shown in Fig. 3(a) illustrates the topography of S. mutans cells with different heights, while (b) shows the height profiles of S.mutans cells in red and culturing medium in green. Figure. 3(c) depicts the height distribution of S. mutans and culturing medium on glass substrate. The Figure revealed that S. mutans cells are of height from 250nm to 500nm with maximum count at 350nm, and a few stacked-up cells have heights of 600 nm to 650nm. These heights correspond to the bright spots in Fig. 3(a). The height distribution of dried culturing medium (BHI and sucrose) on glass substrates in Fig. 3(c) shows height ranging from 10nm to 200nm with maximum count at 80nm.

*3.2   MIE-RCE Results*

Before performing ellipsometry measurements of S. mutans cells on glass substrate, the $\Psi$ and $\Delta$ spectra were measured for a bare glass substrate with a carbon tape attached to the backside in order to block the backside reflections of the glass substrate. Cauchy model, [35] was used to fit $\Psi$ and $\Delta$ spectra in the visible range (450nm - 750nm) and obtain the refractive index of glass substrate (1.549, 1.538, 1.532 at 500nm, 600nm, and 700nm respectively). MIE measurements of S. mutans samples cultured on a glass substrate for 72 hours were done using specular mode at an angle of incidence (AOI) of 52º in the visible range (450nm – 750nm) and the analyses were done for three specific wavelengths. In order to overcome the signal noise in ellipsometry measurements, the values of $\Psi$ and $\Delta$ of 5×5 pixels are averaged to produce a mean value which is registered for the center pixel. We shift the center pixel by 2 pixels (in both x and y directions) and then taking average of the 5×5 pixels around it, and then move on. This leads to smoothed image with a total area of 58 grids×58 grids, with each grid being 328nm×328nm in size (which contains signals average over the surrounding 820nm× 820nm area). The averaging procedure was done by using an in-house MATLAB program to produce smoothed images of the sample.

To show the difference among single-bounce, two-bounce- and multi-bounce analyses, $\tan\Psi$ and $\cos\Delta$ obtained for the three models are plotted as functions of $n_1$ in the interval (1, 3) as shown in Fig. 4. Here, $\tan \Psi = |r_p/r_s|$ and $\cos \Delta = \text{Re}(r_p/r_s)/\tan\Psi$. In Fig. 4 $\tan \Psi$ and $\cos \Delta$ are plotted for three different wavelengths 500nm, 600nm, and 700nm, except for the single-bounce case in Fig. 4(a) since the result is independent of wavelength. The height of the sample considered in Fig. 4(a) and (b) was set to 80nm and 350nm, where 80nm is the average height of dried culturing medium and 350nm is the average height of S.mutans revealed by AFM measurements in Fig. 3(c).

Figure 4(a) and (b) show that results obtained by two-bounce and multi-bounce analyses overlap well for two different heights. This means that results of samples analyzed by two-bounce and multi-bounce models will be very close. As a calibration, in Fig. 5 we show the results of $\Psi$ and $\Delta$ images for culturing medium sample (without S.mutans) at wavelength equal to 600nm analyzed using single-bounce, two-bounce, and multi-bounce models.

We noticed that the topology of the $n_1$ image is very similar to the $\Psi$ image, while the $k_1$ image is very similar to the $\Delta$ image. Comparing $n_1$ images with the corresponding $\Psi$ images in Fig. 5 we conclude that the cyan color areas in $n_1$ images with $n_1$ values ranging from 1.25 (1.39) to 1.3 (1.41) in single-bounce (two-bounce and multi-bounce) analyses coincide with the areas in dark blue of $\Psi$ images. The yellow areas of $n_1$ images match the light blue areas in $\Psi$ images, while the dark blue spots in $n_1$ images match the red spots in $\Psi$ images. In single-bounce analysis the yellow-color areas in the $n_1$ image corresponds to the blue areas in the $k_1$ image which has negative (unphysical) values. This could be caused by depolarization of scattered light or it's an indication that those areas are flat and not suitable for single-bounce analysis. Even with unphysical $k_1$ values in some areas, the $n_1$ values obtained by the single-bounce model are still fairly close to the results obtained by two-bounce and multi-bounce methods. The blue spots in the $n_1$ image correspond to the yellow and red spots in the $k_1$ image which has positive values, indicating possible absorption there.

The $n_1$ images obtained from two-bounce and multi-bounce analyses are very close (with difference less than 0.02 in all pixels). In the multi-bounce analysis there are two roots for $n_1$ corresponding to the "+" and "-" signs in Eq. (A11). It is interesting to note that the physical results for $n_1$ of the culturing medium are found only by adopting the "-" sign in Eq. (A11) for this sample.

Comparing the $k_1$ image in single-bounce analysis with $h$ images in two-bounce and multi-bounce analyses, we found that the blue stripes in the $k_1$ image match the red stripes in $h$ images with height between 200nm and 280nm. Other areas correspond to $k_1$ less than zero and height less than 40nm, which are undetectable by single-bounce analysis. It is interesting to note that the height of the culturing medium revealed by the two-bounce and multi-bounce analyses is either below 40nm (flat areas) or above 200nm (bumpy areas) and the topography of the $h$ image is very close to the $k_1$ image and $\Delta$ image.

*3.2.1 Single-Bounce analysis*

Single-bounce model was first used to analyze S. mutans with culturing medium on glass substrate. The left side images of Figs. 6 and 7 show the CCD images of two measured areas (area 1 and area 2) of interest for S. mutans samples. The measured $\Psi$ and $\Delta$ images at wavelengths of 500nm, 600nm, and 700nm for the two different areas are shown in the second and third columns of Figs. 6 and 7.

$\Psi$ and $\Delta$ images at different wavelengths were used to calculate $\rho$ in Eq. (2). The obtained value of $\rho$ is substituted in Eq. (3), and our in-house MATLAB program was used to solve this equation and find values of refractive index $n_1$ and extinction coefficient $k_1$ for the three wavelengths and show their images in the fourth and fifth columns of Figs. 6 and 7. We noticed again that the topology of the $n_1$ image is very similar to the $\Psi$ image, while the $k_1$ image is very similar to the $\Delta$ image. Comparing $n_1$ images with the corresponding $\Psi$ images in Figs. 6 and 7 for the three wavelengths we conclude that areas in blue color of $n_1$ images with $n_1$ ranging from 1.2 to 1.4 with a few (dark blue) spots less than 1.2 coincide with the blue areas in $\Psi$ images.

The cyan area and green stripes (green and yellow stripes) in Fig. 6 (Fig.7) for $k_1$ images have positive values between 0 and 0.2 which are physically reasonable, and they correspond to the blue areas in $\Delta$ images with values between -180º and -160º. The corresponding $n_1$ values for these areas are between 1.6 and 2 which correspond to bumpy areas in culturing medium. There are a few spots in $k_1$ images with values larger than 0.2 which indicates strong light trapping of the bumpy structure. The areas in $k_1$ images with negative values (blue to cyan) are attributed to strong depolarization effect and the corresponding values of $n_1$ there are between 1.2 and 1.6.

It should be noted that the depolarization effect due to scattering from S. mutans cells and rough areas of culturing medium has been neglected in this model. Thus, the $n_1$ and $k_1$ values obtained can be rather crude. The $k_1$ value obtained can be misleading, since it is very sensitive to the phase and the depolarization effect. On the other hand, this one-bounce analysis is quite effective in locating which area is bumpy and the distribution of S. mutans cells can be related to the $n_1$ distribution obtained.

*3.2.2 Two-Bounce analysis*

For the two-bounce analysis, the calculated $\rho$ from measured $\Psi$ and $\Delta$ images were substituted in Eq. (6). This equation was solved for different wavelengths to find the refractive index. The third column of Figs. 8 and 9 show the $n_1$ images for the two samples obtained by solving Eq. (6) for wavelengths of 500nm, 600nm, and 700nm.

Areas in cyan for $n_1$ images in Figs. 8 and 9 show $n_1$ values ranging from 1.2 to 1.4, and areas with light green and some grids with yellow, orange and red colors show values ranging from 1.4 to 2.6. Comparing results of $n_1$ images obtained from two-bounce and single-bounce analysis, we found that areas in cyan in Figs. 8 and 9 correspond to same areas in blue color in Figs. 6 and 7, and $n_1$ values obtained by both models are similar, although the $k_1$ values there are negative (unphysical). This indicates that bacteria cells surrounded by culturing medium represented by the blue areas in $n_1$ images in Figs. 6 and 7 can still be detected by single-bounce analysis, and the unphysical values of $k_1$ there are due to depolarization effect. This is confirmed by the two-bounce analysis.

The two-bounce model can be used to analyze and sense the S.mutans cells in areas where the height distribution of cells is uniform, similar to height profile at 350nm as shown in AFM results (green line) of Fig. 3(b). In addition, the refractive index values obtained from the two-bounce analysis can be used to calculate the height distribution of S.mutans cells surrounded by culturing medium according to Eq. (9). Thus, the two-bounce analysis is more versatile for the current application. The fifth columns of Figs. 8 and 9 show the obtained height (*h*) distribution for area 1 and area 2. The blue and cyan areas with *h* ranging from 40nm to 200nm in Figs. 8 and 9 correspond to areas filled with culturing medium. Areas with *h* above 200nm (green, yellow, orange, and red) correspond to areas filled with cells, in agreement with the height distribution of S. mutans cells and dried culturing medium obtained by AFM in Fig. 3(c).

*3.2.3 Multi-Bounce analysis*

The first two columns of Fig. 10 show images of $n_1$ and $h$ distributions *in* area 1 for $\lambda=500$nm obtained by solving Eq. (8) with both "-" and "+" signs in Eq. (A11), respectively. In both calculations, the dark blue area represents the unphysical zone (since $n_1$ obtained there is less than 1). We found that the two solutions are mutually exclusive with the unphysical zone obtained with "-" sign being the physical zone obtained with "+" sign, and vice versa. Furthermore, the physical values of $n_1$ range from 1.3 to 1.5 (red zone) in the first column and from 1.5 to 3 (mixed-color zone) in the second column. The corresponding heights in the first column ranges from 200nm to 300nm, while in the second column it ranges from 20nm to 180nm. So, the two solutions of the multi-bounce analysis can be used to distinguish regions of low index (high $h$ value) and high index (low $h$ value), which correspond to regions filled with high and low concentrations of S. Mutans cells, respectively. We can combine the physical solutions for $n_1$ obtained with both signs, and the outcome is presented in the third column. The result is very similar to that obtained by the two-bounce analysis as shown in the first row of Fig. 8.

In the above analysis we have neglected the effects of depolarization and stray light, which can lead to errors in our estimate of $n_1$ and $h$. Thus, the $n_1$ and height distributions deduced can be different for different wavelengths as shown in Figs. 6 and 7 as well as in Figs. 8 and 9. The height distribution analysis for different wavelengths of S.mutans and dried culturing medium in our study should be wavelength independent and the variation of $n_1$ in the range of wavelengths considered should also be negligible. Thus, this variation in results for different wavelengths gives an estimate of the error caused by depolarization effect. The error estimate of $n_1$ values obtained by single bounce and two bounce (same as multi-bounce) analyses are reported in Table 1, which shows a standard deviation in the 1% range. For the $h$ distribution, since $h$ varies significantly from pixel to pixel in the samples considered and the image resolution depends on wavelength, it is not straightforward to estimate the error in $h$ distribution by comparing results for different wavelengths. However, we can see that in areas where the height distribution is uniform, the results obtained become rather insensitive to the wavelength, while in bumpy areas the variation can be significant. The overall ranges and the topologies of $h$ in Figs. 8 and 9 for different wavelengths are also similar. However, changes of distributions for wavelength from 500nm to 700nm are clearly visible.

Finally, we give some estimate of the statistical errors in our measurements. We determine the average standard deviation ($\bar{\sigma}$) by repeating the same measurements five times for MIE measurements of a bare glass substrate with carbon tape attached to the backside. The mean value of $\Psi$ and $\Delta$ maps ($\bar{\Psi}, \bar{\Delta}$) and average standard deviation ($\bar{\sigma}$) of this study are shown in Table. 2. These statistical values were calculated for single-pixel maps and maps obtained by averaging over 5 surrounding pixels, since results reported in this study were based on $\Psi$ and $\Delta$ maps averaged over 5 surrounding pixels explained in Sec. 3.2. As shown in Table 2, the average standard deviation ($\bar{\sigma}$) for the 5-pixel average map is about one half of the single-pixel map, thus the resulting MIE image will be less noisy.

We can also estimate the error introduced in our $\Psi$ and $\Delta$ values determined by using finite number of compensator angles ($\phi_c$) in our measurements. If there is no statistical error or systematic error, one should be able to determine the five Fourier coefficients in Eq. (1) with just five values of $\phi_c$. Given the measurements at 36 values of $\phi_c$, we can select them all (with 10º spacing), or adopt only 18 values at 20º spacing, or 12 values at 30º spacing. We then have three ways to determine the $\Psi$ and $\Delta$ values. The deviation of these determined values from the mean gives us an estimate of the measurement error. The estimated average standard deviation $\bar{\sigma}$ for areas 1 and 2 of our sample with S. mutans cells are also listed in Table 2.

For single-pixel $\Psi$ maps, error ranges from 1.8% for glass substrate to 4.7% for area 1, and 5.9% for area 2. In single-pixel $\Delta$ maps error ranges from 0.3% for glass substrate to 1.4% for area1, and 1.3% for area 2, indicating that 5-pixel averaged maps are more accurate and reproducible by MIE.

## 4    Conclusions

In this paper, we performed MIE measurements in the visible range to detect dried S. mutans cells and surrounding culturing medium on glass substrate. We found that two-bounce and multi-bounce lead to very close results for refractive index values and height distributions. Incorporating single-bounce, two-bounce, and multi-bounce analyses can reveal more information about the biological sample under investigation. The two-bounce and multi-bounce models provide the height distribution, while the single-bounce model provides additional information on extinction coefficient or the depolarization effect. Results obtained in this work show the ability of MIE to detect biological structures and extract their average refractive index and height distribution. One of the major motivations of using MIE in biology is to gain more knowledge about the dielectric properties of biological systems for better understanding and further development of biomaterials. Moreover, it constitutes important driving forces in developments directed towards sensors and diagnostic applications for use in clinic medicine.[4] Results obtained in this work show that MIE is a valuable detecting technique, which can differentiate bacteria types by examining their morphologies and dielectric properties. Since it's a label-free, non-invasive sensing technique, it does not require any sample preparation, and it can perform measurements within seconds. Because of these advantages, MIE allows studying the details about the interaction mechanism between a substrate and adsorbed layer for samples under investigation via real-time measurements and analysis. The amount of information obtained from the analysis of MIE measurements can be extended to include the effects due to surface roughness and molecular orientation.[4] Finally, comparing this sensing tool to other label-free and non-invasive imaging techniques such as Raman spectroscopy and IR/thermal imaging, MIE provides complementary information of the samples compared to Raman spectroscopy[36] and IR/thermal imaging.[37] Since Raman spectroscopy measures the vibrational modes of the sample, it is more sensitive to the chemical nature of the sample, while MIE is capable of detecting the distribution of cells and measuring their thickness and dielectric properties. The IR/thermal imaging technique detects the thermal radiation from the sample, which cannot achieve the submicron spatial resolution considered here, and it does not have the capability of extracting detailed structure information as MIE does. Although we only report MIE technique in the visible range, the same experimental setup and analysis can be applicable for other wavelengths such as ultraviolet or infrared.

# APPENDIX A: DERIVATION OF FORMULAS FOR THE THREE MODELS USED

## A.1 Single-Bounce light Path

The Fresnel reflection coefficients $r_p$ and $r_s$ in z-direction for single-bounce light path from Fig. 2(a) are:

$$r_{01p} = \frac{E_{rx}}{E_{ix}} = \frac{\frac{1}{\varepsilon_0}K_{iz} - \frac{1}{\varepsilon_1}K_{tz}}{\frac{1}{\varepsilon_0}K_{iz} + \frac{1}{\varepsilon_1}K_{tz}} = \frac{\frac{1}{\varepsilon_0}\sqrt{\varepsilon_0 - \sin\theta_0^2} - \frac{1}{\varepsilon_1}\sqrt{\varepsilon_1 - \sin\theta_0^2}}{\frac{1}{\varepsilon_0}\sqrt{\varepsilon_0 - \sin\theta_0^2} + \frac{1}{\varepsilon_1}\sqrt{\varepsilon_1 - \sin\theta_0^2}} \quad , \tag{A1}$$

$$r_{01s} = \frac{E_{ry}}{E_{iy}} = \frac{K_{0z} - K_{1z}}{K_{0z} + K_{1z}} = \frac{\sqrt{\varepsilon_0 - \sin\theta_0^2} - \sqrt{\varepsilon_1 - \sin\theta_0^2}}{\sqrt{\varepsilon_0 - \sin\theta_0^2} + \sqrt{\varepsilon_1 - \sin\theta_0^2}} \quad , \tag{A2}$$

where $K_{iz}$ ($i = 0,1$) denotes the z-component of the wave-vector for light in media $i$. $\theta_0$, $\varepsilon_0$, and $\varepsilon_1$ are the angle of incidence, the complex refractive index of air, and the complex refractive index of S. mutans cells, respectively. Substituting for $r_p$ and $r_s$ in Eq. (2) in Section 2.4 we obtain,

$$\rho = \tan\Psi \; e^{i\Delta} = \frac{\left(\frac{1}{\varepsilon_0}\sqrt{\varepsilon_0 - \sin\theta_0^2} - \frac{1}{\varepsilon_1}\sqrt{\varepsilon_1 - \sin\theta_0^2}\right)\left(\sqrt{\varepsilon_0 - \sin\theta_0^2} + \sqrt{\varepsilon_1 - \sin\theta_0^2}\right)}{\left(\frac{1}{\varepsilon_0}\sqrt{\varepsilon_0 - \sin\theta_0^2} + \frac{1}{\varepsilon_1}\sqrt{\varepsilon_1 - \sin\theta_0^2}\right)\left(\sqrt{\varepsilon_0 - \sin\theta_0^2} - \sqrt{\varepsilon_1 - \sin\theta_0^2}\right)} \quad , \tag{A3}$$

Solving Eq. (A3) with respect to $\varepsilon_1$, we have

$$\varepsilon_1 = (n_1 + ik_1)^2 = \frac{(1-\rho)^2}{(1+\rho)^2}\sin\theta_0^2 \tan\theta_0^2 + \sin\theta_0^2 \quad , \tag{A4}$$

## A.2 Two-Bounce light Path

For the two-bounce light path depicted in Fig. 2(b) the Fresnel reflection coefficients is given by:

$$r_j = r_{01j} + \left(1 - r_{01j}^2\right) r_{12j} \; e^{2i\beta} \quad , \tag{A5}$$

where *j* stands for s-polarization and p-polarization, and β represents the phase film factor given by the equation below:

$$\beta = \frac{2\pi h n_1 \cos\theta_1}{\lambda} ,$$ (A6)

where *h* and $n_1$ are the thickness and refractive index of the sample.

Substituting Eq. (A5) for *s* and *p* polarization in Eq. (2), Section 2.4, we obtain

$$\rho = \tan\Psi \; e^{i\Delta} = \frac{r_{01p} + (1 - r_{01p}^2) r_{12p} \gamma}{r_{01s} + (1 - r_{01s}^2) r_{12s} \gamma} ,$$ (A7)

where $r_{01p}$, $r_{01s}$, $r_{12p}$, and $r_{12s}$ have been derived in a previous work by Taya et al.[30]

Solving Eq. (A7) with respect to $e^{-2i\beta}$ we obtain

$$\gamma = e^{2i\beta} = \frac{r_{01p} - \rho r_{01s}}{\rho (1 - r_{01s}^2) r_{12s} - (1 - r_{01p}^2) r_{12p}} ,$$ (A8)

Assuming the sample does not absorb light in the visible region (*i.e.*, $k_1 = 0$), then

$$\ln|\gamma| = \ln \left| \frac{r_{01p} - \rho r_{01s}}{\rho (1 - r_{01s}^2) r_{12s} - (1 - r_{01p}^2) r_{12p}} \right| = 0$$ (A9)

### A.3 Multi-Bounce light Path

The Fresnel reflection coefficients of multi-bounce light paths have been derived in Ref.[32] and given as,

$$\rho = \tan\Psi \; e^{i\Delta} = \frac{(r_{01p} + r_{12p} \gamma)(1 + r_{01s} r_{12s} \gamma)}{(r_{01s} + r_{12s} \gamma)(1 + r_{01p} r_{12p} \gamma)} ,$$ (A10)

Assuming that $k_1 = 0$ in the visible range then the final solution is written as:

$$\ln|\gamma| = \ln|e^{2i\beta}| = \ln\left|\frac{-B \pm \sqrt{B^2 - 4AC}}{2A}\right| = 0 ,$$ (A11)

where

$$A = (\rho \, r_{01p} - r_{01s})(r_{12p} \, r_{12s}) ,$$ (A12)

$$B = r_{01p}r_{01s}\left(\rho\, r_{12p} - r_{12s}\right) + \rho\, r_{12s} - r_{12p}\ , \tag{A13}$$

$$C = \rho\, r_{01s} - r_{01p}\ , \tag{A14}$$

The thickness of the sample can be obtained from Eq. (A6) in terms of $\beta$, $n_1$, and $\cos\theta_1$ as:

$$h = \frac{\beta\, \lambda}{2\,\pi\, n_1\, \cos\theta_1} \tag{A15}$$

*Disclosures*

No conflicts of interest, financial or otherwise, are declared by the authors.

*Acknowledgments*

This work was supported in part by the Ministry of Science and Technology of Taiwan under Contract No. MOST 104-2112-M-001-009-MY2.

**Mai Ibrahim Khaleel** is a PhD student in the nanoscience and technology program-Taiwan International Graduate Program - in a collaboration with the department of engineering and system science at National Tsing Hua University (NTHU). She received her BS and MS degrees in physics from Al-Quds University in 2007 and 2009, respectively. Her current research interests include bio-optics, imaging ellipsometry, and photonics.  email: maik@gate.sinica.edu.tw

**Figures and Tables**

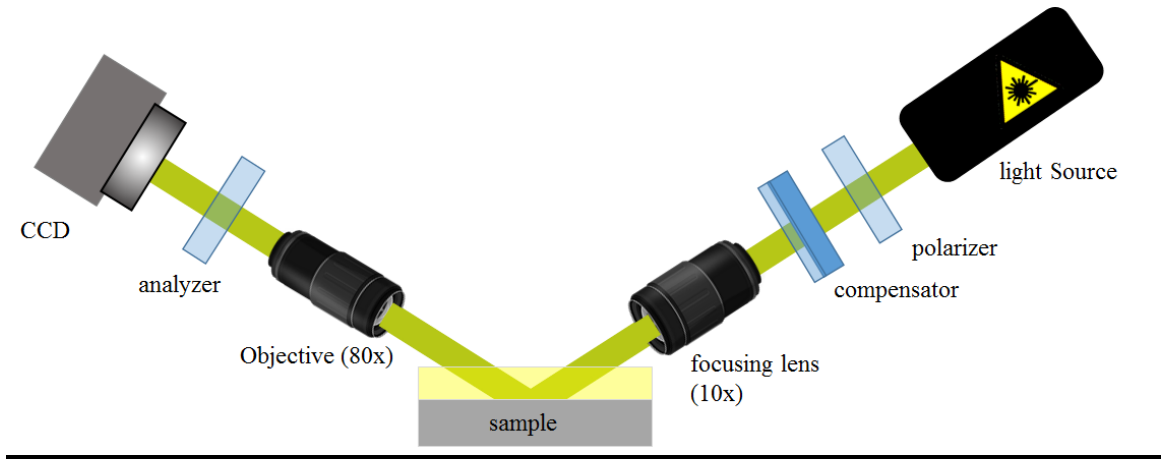

**Fig. 1** Schematic diagram for rotating compensator ellipsometry.

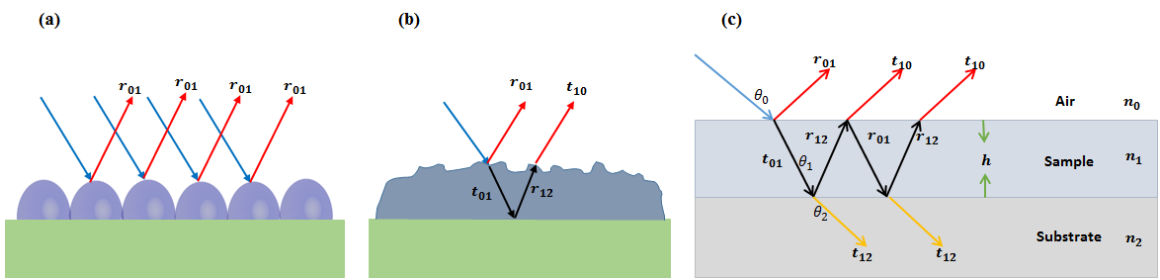

**Fig. 2** Schematic diagram for light scattering paths from the sample under consideration: (a) single-bounce detection, (b) two-bounce detection, and (c) multi-bounce detection.

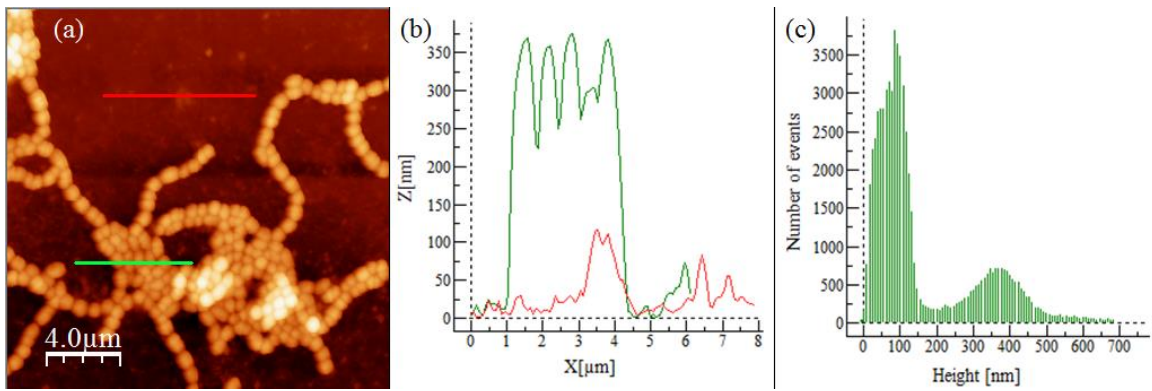

**Fig. 3** S. mutans on a glass substrate after 72-hours culture: (a) 20 μm × 20 μm AFM topography image, (b) height profile of culturing medium (red) and S.mutans cells (green), and (c) height distribution of culturing medium and S. mutans cells.

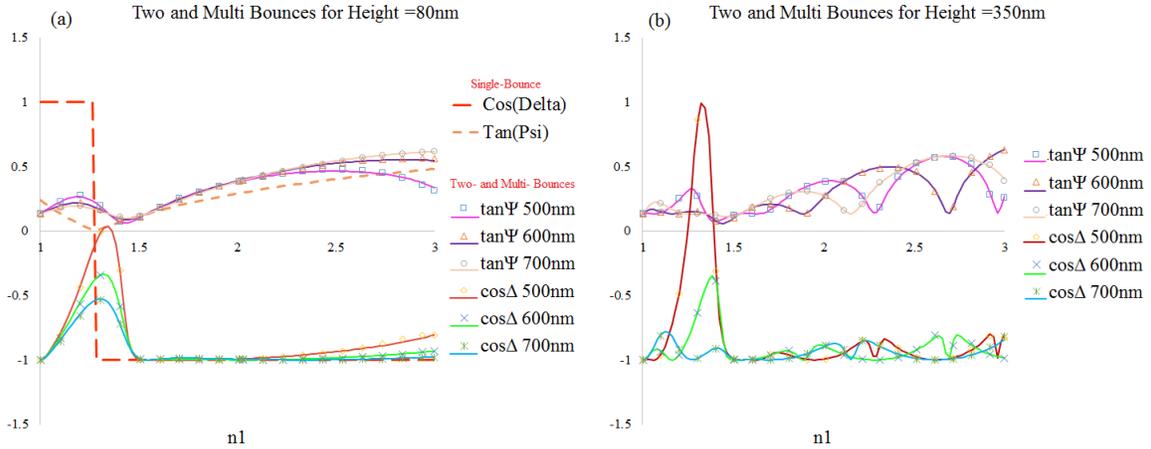

**Fig. 4** tan$\Psi$ and cos$\Delta$ as functions of $n_1$ at AOI of 52° for (a) two-bounce model (symbols), and multi-bounce model (continuous curves) at height = 80nm, and (b) two-bounce model (symbols), and multi-bounce model (continuous curves) at height = 350nm. Single-bounce results are also included in (a) to compare with two- and multi-bounce models.

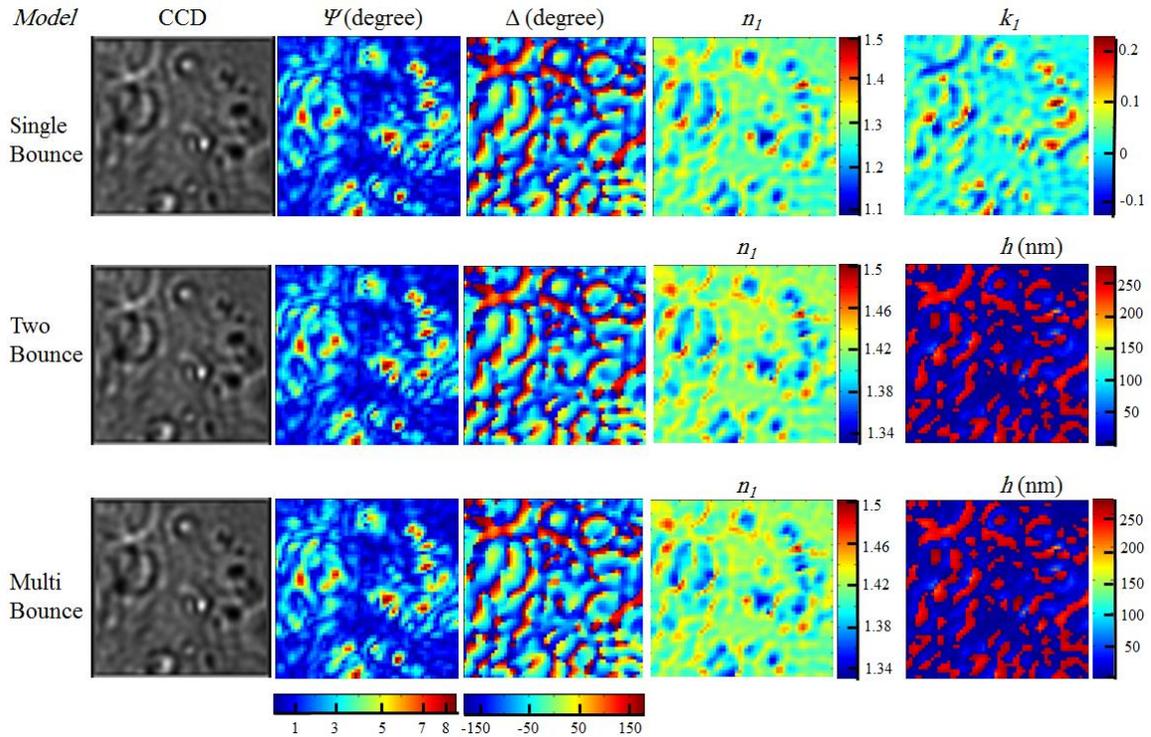

**Fig. 5** CCD, $\Psi$, $\Delta$, $n_1$, $k_1$, and $h$ images of BHI sample using single-bounce, two-bounce, and multi bounce models at AOI=52°. The frame size is 19 μm × 19 μm. The units of $\Psi$ and $\Delta$ are degrees.

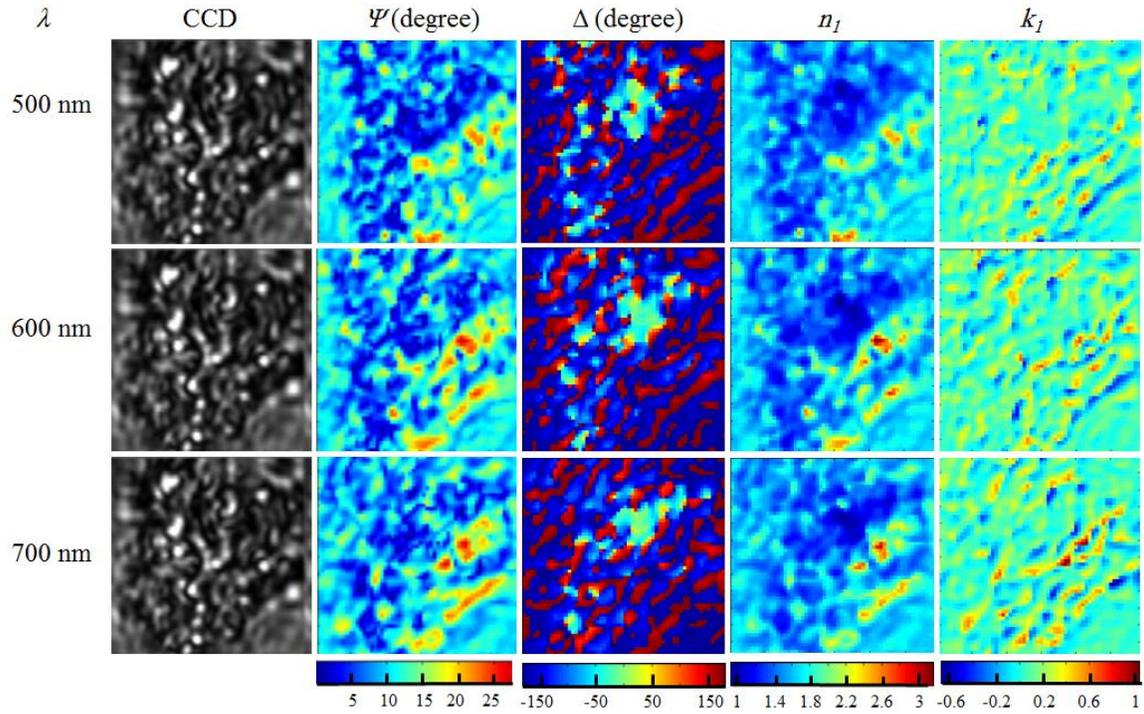

**Fig. 6** CCD, $\Psi$, $\Delta$, refractive index $n_1$, and extinction coefficient $k_1$ images of S. mutans sample for area 1 using single-bounce analysis at AOI=52°. The frame size is 19 μm × 19 μm. The units of $\Psi$ and $\Delta$ are degrees.

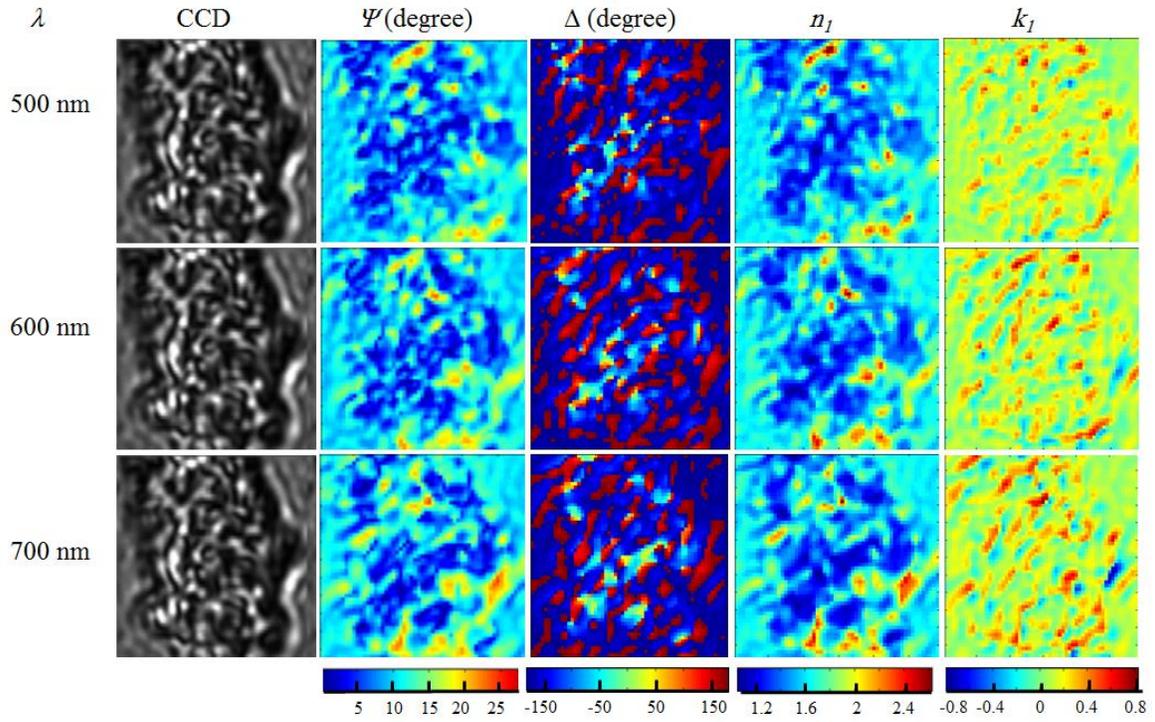

**Fig. 7** CCD, $\Psi$, $\Delta$, refractive index $n_1$, and extinction coefficient $k_1$ images of S. mutans sample for area 2 using single-bounce analysis at AOI=52º. The frame size is 19 μm × 19 μm. The units of $\Psi$ and $\Delta$ are degrees.

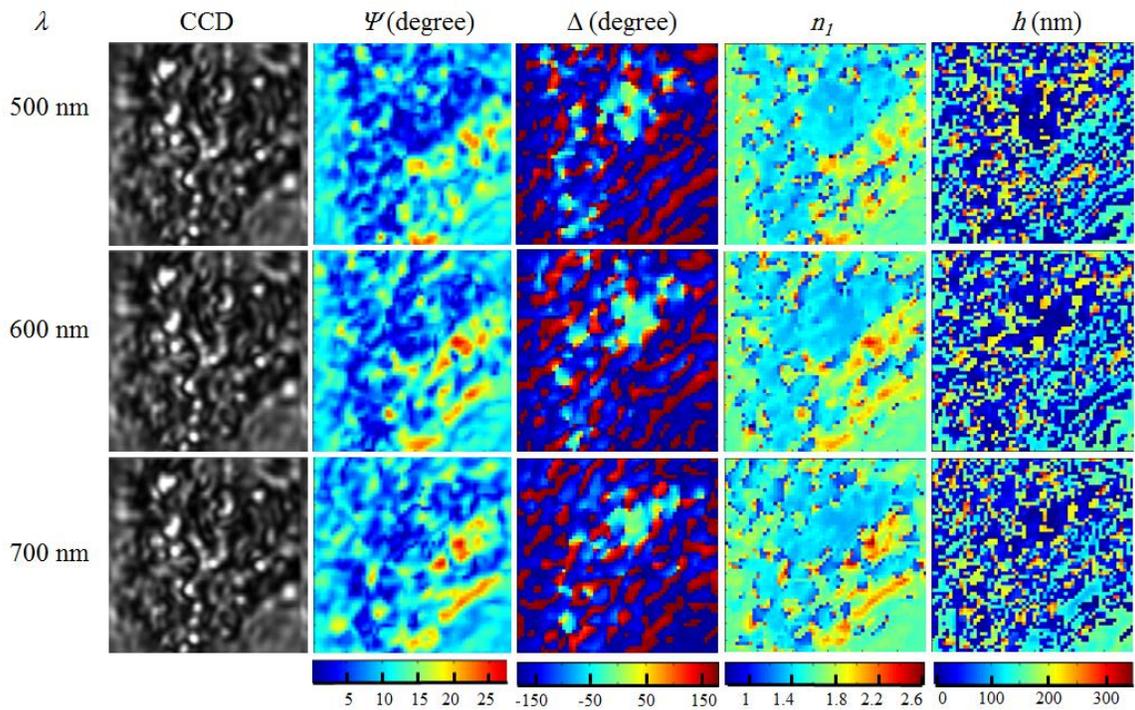

**Fig.8** CCD, $\Psi$, $\Delta$, refractive index $n_1$, and height distribution $h$ images of S. mutans sample for area 1 using two-bounce analysis at AOI=52°. The frame size is 19 µm × 19 µm. The units of $\Psi$ and $\Delta$ are degrees.

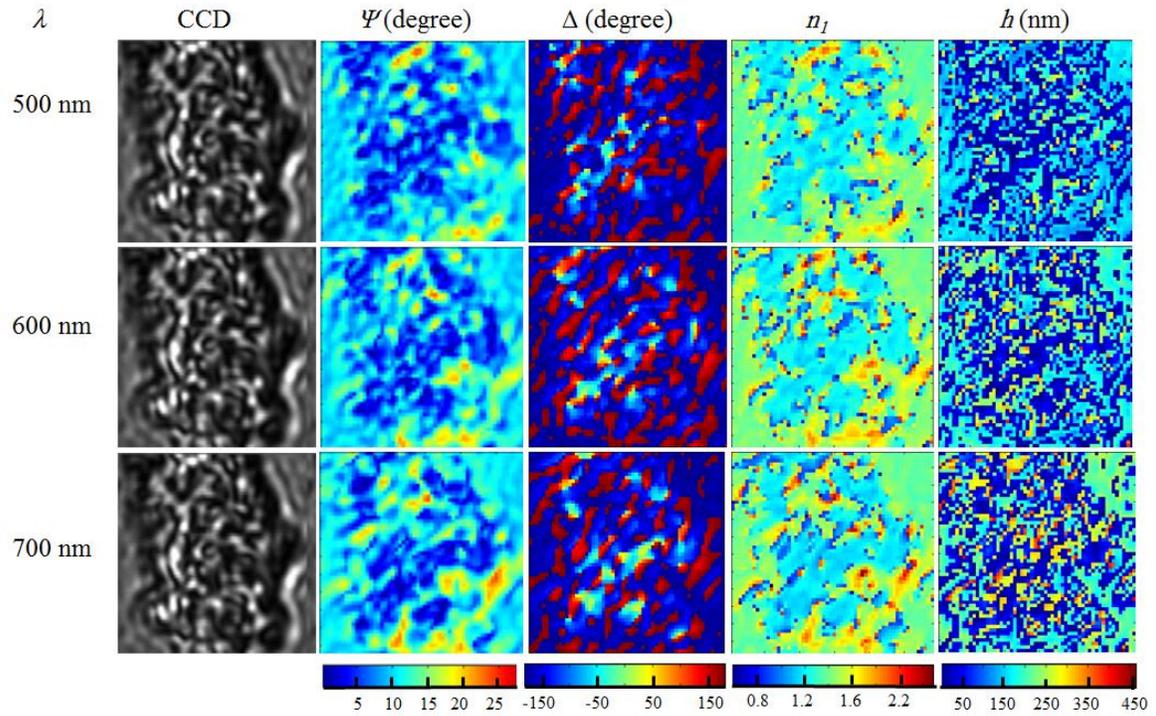

**Fig.9** CCD, $\Psi$, $\Delta$, refractive index $n_1$, and height distribution $h$ images of S. mutans sample for area 2 using two-bounce analysis at AOI=52°. The frame size is 19 µm × 19 µm. The units of $\Psi$ and $\Delta$ are degrees.

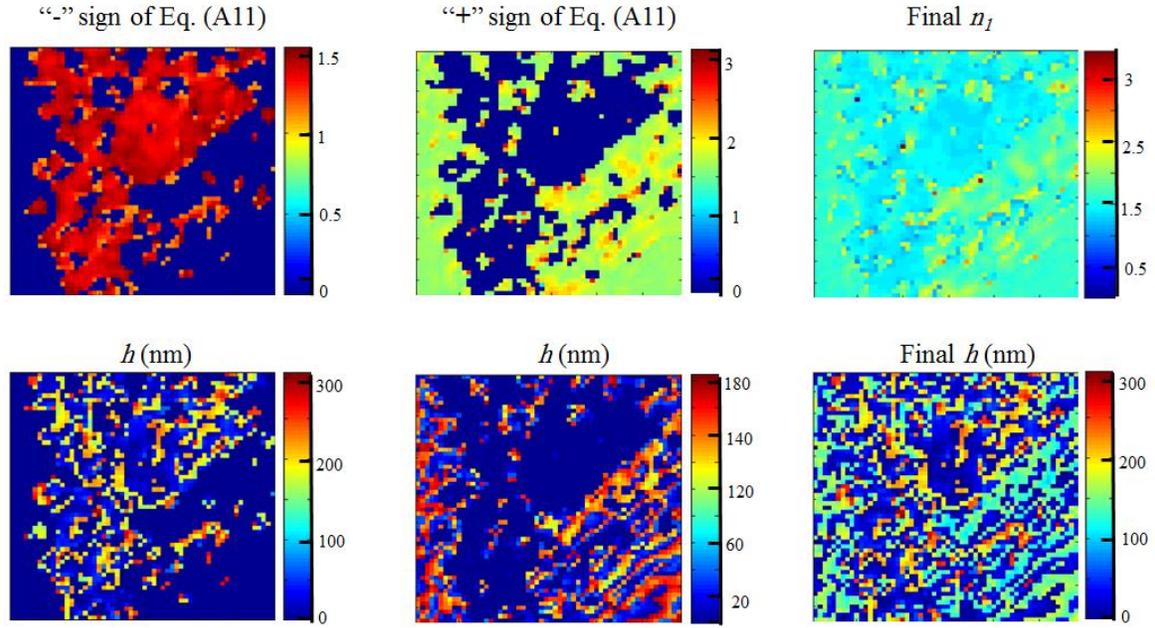

**Fig. 10** Multi-bounce analysis for refractive index $n_1$ obtained by using "+" and "-" signs in Eq. (A11) and height ($h$) distribution images of S. mutans sample for area 1 at λ=500nm at AOI=52º. The frame size is 19 μm × 19 μm. The units of $\Psi$ and $\Delta$ are degrees.

**Table. 1** Standard deviation and mean value of $n_1$ obtained by single bounce and two bounce models for area 1 and area 2 at AOI=52º.

|  | Single bounce $n_1$ | | Two bounce $n_1$ | |
| --- | --- | --- | --- | --- |
|  | $\bar{\sigma}$ | $\overline{n_1}$ | $\bar{\sigma}$ | $\overline{n_1}$ |
| Area1 | 0.144 | 1.576 | 0.169 | 1.557 |
| Area2 | 0.13 | 1.59 | 0.161 | 1.567 |

**Table 2** Standard deviation and mean value of Ψ and Δ measured by MIE for glass substrate, area 1 and area 2 at AOI=52º.

|  |  | Glass substrate | | | | Area1 | | | | Area2 | | | |
|---|---|---|---|---|---|---|---|---|---|---|---|---|---|
|  |  | Ψ | | Δ | | Ψ | | Δ | | Ψ | | Δ | |
|  |  | $\bar{\sigma}$ | $\bar{\Psi}$ | $\bar{\sigma}$ | $\overline{|\Delta|}$ | $\bar{\sigma}$ | $\bar{\Psi}$ | $\bar{\sigma}$ | $\overline{|\Delta|}$ | $\bar{\sigma}$ | $\bar{\Psi}$ | $\bar{\sigma}$ | $\overline{|\Delta|}$ |
| 500nm | Single pixel | 0.155 | 14.62 | 0.653 | 178.2 | 0.33 | 9.825 | 2.314 | 167.2 | 0.3 | 9.169 | 2.07 | 175.9 |
| | 5-pixel average | 0.111 | 14.71 | 0.524 | 178.9 | 0.131 | 10.32 | 0.993 | 178.1 | 0.113 | 8.102 | 0.726 | 173.1 |
| 600nm | Single pixel | 0.251 | 14.32 | 0.563 | 178.8 | 0.346 | 10.77 | 2.1 | 168.9 | 0.337 | 8.3 | 2.03 | 168.1 |
| | 5-pixel average | 0.15 | 14.36 | 0.468 | 179.1 | 0.106 | 10.33 | 0.708 | 174.5 | 0.107 | 8.659 | 0.646 | 177 |
| 700nm | Single pixel | 0.269 | 14.77 | 0.64 | 178.8 | 0.476 | 9.951 | 2.533 | 177.6 | 0.442 | 7.481 | 2.377 | 178.3 |
| | 5-pixel average | 0.127 | 14.45 | 0.526 | 179 | 0.157 | 10.34 | 0.765 | 173 | 0.134 | 10.42 | 0.78 | 174 |